# Influence of projection effects on the observed differential rotation rate in the UV corona

Salvatore Mancuso, Silvio Giordano

*INAF - Osservatorio Astronomico di Torino, ITALY*

**Abstract**

Following previous investigations by Giordano and Mancuso [1] and Mancuso and Giordano [2, 3] on the differential rotation of the solar corona as obtained through the analysis of the intensity time series of the O VI 1032 Å spectral line observed by the UVCS/SOHO telescope during solar cycle 23, we analysed the possible influence of projection effects of extended coronal structures on the observed differential rotation rate in the ultraviolet corona. Through a simple geometrical model, we found that, especially at higher latitudes, the differential rotation may be less rigid than observed, since features at higher latitudes could be actually linked to much lower coronal structures due to projection effects. At solar maximum, the latitudinal rigidity of the UV corona, with respect to the differential rotating photosphere, has thus to be considered as an upper limit of the possible rigidity. At solar minimum and near the equatorial region throughout the solar cycle, projection effects are negligible.

# Introduction

The coronal rotation rate and its variation as a function of height and heliographic latitude remains as yet a poorly understood and debated topic. In fact, quantifying the coronal rotation rate critically depends upon both the applied methods of data reduction and the type of analysed data. The rotation rate in the solar corona has been studied by means of different coronal tracers over an extended set of wavelengths, covering the range from radio to X-rays. In the past few years, the analysis of a full decade of data obtained by telescopes aboard the Solar and Heliospheric Observatory (SOHO; [4]) spacecraft, has allowed to study the temporal variation in the coronal rotation rate for the whole solar cycle 23. In particular, time-series observations of the coronal O VI 1032 Å spectral line intensity provided by the UltraViolet Coronagraph Spectrometer (UVCS/SOHO; [5]) telescope on board SOHO have been effectively used to study the differential rotation of the solar corona during minimum [1] and maximum [2] solar activity and throughout the whole solar cycle [3]. The work of Giordano & Mancuso [1] confirmed the already established result that the corona, during minimum activity, tends to rotate with a less pronounced differential rotation than the plasma of the photosphere. The estimated equatorial synodic rotation period of the corona was ~27.5 days. Mancuso & Giordano [2] in a similar study carried out during a four-year period around solar maximum, showed that the coronal rotation differential profile tends to be less rigid, that is, more differential, during enhanced solar activity. In general, during solar maximum, the coronal magnetic structures were observed to rotate much faster at all latitudes, and less differentially, than the underlying small-scale magnetic structures linked to the photospheric plasma. A striking significant positive correlation was finally discovered by Mancuso & Giordano [3] between the variations in the residual rotation rates of the coronal and sub-photospheric equatorial plasma, suggesting that the observed variations in the coronal rotation rate reflect the dynamic changes inferred within the near-surface shear layer, where the tracer structures responsible for the observed coronal emission are thus most probably anchored. Projection effects could certainly play a major role in the interpretation of the results presented in the previous works. In this paper, we analyse the possible influence of projection effects of extended coronal structures on the observed differential rotation rate in the ultraviolet corona.

## Data reduction and analysis

The data analyzed by Giordano and Mancuso [1] and Mancuso and Giordano [2, 3] were collected in a time interval from April 1996 to May 2007 from observations of the coronal O VI 1032 Å spectral line, which is routinely observed by the UVCS/SOHO instrument. UVCS is an internally and externally occulted coronagraph consisting of two spectrometric channels for the observation of spectral lines in the UV range and a visible light channel for polarimetric measurements of the extended solar corona. The UVCS slit, parallel to a tangent to the solar limb on the plane of the sky, can be moved along the radial direction, thus being able to yield raster observations of the solar corona between 1.4 and 10 $R_\odot$ with a field of view of 40'. To cover all possible position angles, the slit can be rotated by 360° about an axis pointing to the Sun's center. For a complete description of the UVCS instrument, see [5]. The periodicity analysis was restricted to periods on time scales near the 27-day solar rotation period and was obtained by combining results from the east and west hemispheres. Fig. 1 shows an O VI 1032 intensity synoptic map at 1.6 $R_\odot$ in the time interval from March 1999 to December 2002. The lower panels show O VI 1032 intensity synoptic maps for each single year. Indeed, the intensity maps of Fig. 1 show a clear modulation, which can be readily attributed to the rotation of persistent features through several consecutive rotations. Results can be found in papers [1-3].

## Results and discussions

During solar minimum, the global, dipolar-like magnetic field of the Sun is the dominant factor in determining the structure of the UV coronal tracers, strictly linked to a longitudinally modulated streamer belt [6]. However, during the maximum phase, multipolar components become predominant and active regions tend to dominate the magnetic flux up to a factor of three. A major caveat in the analysis proposed in the papers of Giordano and Mancuso [1] and Mancuso and Giordano [2,3] is the possibility that different structures along the line of sight, rooted at different latitudes over the solar surface, might contribute to the observed periodicity due to projection effects of lower latitude features that can contaminate the coronal signal observed at higher latitudes [7]. These projection effects might create an unwanted bias and difficulties in confirming the exact degree of rigidity in the corona. In other words, periodicities apparently observed at higher latitudes might be actually linked to structures from lower latitudes, so that the overall rotation curves could be flatter than observed. We tried to quantify this effect and the bias it may bring to the observed latitude dependence of the coronal rotation rate at 1.6 $R_\odot$. In order to qualitatively evaluate the uncertainty on the latitudinal determination of the position of a bright feature, which acts as a rotation tracer, we used a simple geometric model (see Figure 2) and the empirical determination of the contrast between the tracer and the background corona. In Figure 2, we show a feature (P) lying in the plane defined by the line-of-sight (l.o.s) and the solar rotation axis (z-direction), at the true latitude $\theta_t$ and true heliocentric distance $r_t$. If the brightness of this feature is larger than the background corona, then it can be detected through the integration of the signal along the l.o.s., appearing at the apparent distance $r_a$ and apparent latitude $\theta_a = 0°$. Collecting all the observed O VI 1032 intensity at 1.6 $R_\odot$, we determined the intensity distribution (Fig. 3). The contrast between a bright feature and the background corona was then defined as the ratio between the average plus 3σ intensity and the minimum observed intensity. The so-defined contrast is of about one order of magnitude. On the other hand, we determined the average radial profile of the O VI 1032 intensity from a large number of streamers observed from 1.6 $R_\odot$ to 3.5 $R_\odot$ at solar maximum. We found that the intensity drops out of one order of magnitude from 1.6 $R_\odot$ to 2.3 $R_\odot$. Then, with the assumption that the bright features have a similar profile with height, only those out of the plane of the sky less than 2.3 $R_\odot$ can dominate the emission as they are projected into the plane of the sky to the apparent height of 1.6 $R_\odot$. For a given apparent latitude on the plane of the sky, $\theta_a$, an increasing difference between the apparent and true latitude, $\theta_t$, means an increasing distance from sun center, thus a decreasing of the expected emission. In Fig. 4, for different apparent latitudes, we draw the expected intensity as a function of the true latitude. We expect that when the intensity drops out of one order of magnitude the contamination of the feature out of the plane of the sky is negligible. In Fig. 5, we show the acceptable region of the true latitude as a function of the apparent latitude. In particular, we see that for a tracer observed at about 60° the true latitude ranges from 60° to 68° and for an apparent latitude of 30°, in the worst case, the feature can actually lie at 50°. In this light, at solar maximum, the latitudinal rigidity of the UV corona, with respect to the differential rotating photosphere, has to be considered as an upper limit of the observed rigidity. In fact, the coronal rotation should be more differential if the tracers observed at high latitudes are linked to the lower acceptable latitudes. Although the difficulty in determining the true latitude of a tracer is obviously also present at solar minimum, in that phase of the solar cycle the streamer belt was well defined and restricted to about 25° in latitude from the equator where the projection effects are less relevant.

**Conclusions**

Following previous investigations by Giordano and Mancuso [1] and Mancuso and Giordano [2,3] on the differential rotation of the solar corona as obtained through the analysis of the intensity time series of the O VI 1032 Å spectral line observed by the UVCS/SOHO telescope, we analyzed the possibility that different structures along the line of sight, rooted at different latitudes over the solar surface, might contribute to the observed periodicity due to projection effects. Especially at higher latitudes, the differential rotation may be less rigid than observed, since features observed at higher latitudes could be actually linked to much lower coronal structures due to projection effects. At least during solar maximum and away from the equatorial region, the latitudinal rigidity of the UV corona, with respect to the differential rotating photosphere, has thus to be considered as an upper limit of the possible rigidity.

**Acknowledgements** The authors would like to thank the organizers of IAGA-III symposium, Dr. A. Hady and Dr. L. Damé. The authors used data from the UVCS/SOHO instrument. SOHO is a project of international cooperation between ESA and NASA.

# References

[1] Giordano S, Mancuso S. Coronal rotation at solar minimum from UV observations. Astrophys J 2008;688:656–68

[2] Mancuso S, Giordano S. Differential rotation of the ultraviolet corona at solar maximum. Astrophys J 2011;729:79–86

[3] Mancuso S, Giordano S. Coronal equatorial rotation during solar cycle 23: radial variation and connections with helioseismology. Astron Astrophys 2012;539(A26):1–7

[4] Domingo V, Fleck B, Poland AI. The SOHO mission: an overview. Solar Phys 1995;162:1–37

[5] Kohl JL, Esser R, Gardner LD, Habbal S, Daigneau PS, Dennis EF, et al.. The ultraviolet coronagraph spectrometer for the solar and heliospheric observatory. Solar Phys 1995;162:313–56

[6] Mancuso S, Spangler SR. Faraday rotation and models for the plasma structure of the solar corona. Astrophys J 2000;539:480–91

[7] Lewis DJ, Simnett GM, Brueckner GE, Howard RA, Lamy PL, Schwenn R. LASCO observations of the coronal rotation. Solar Phys 1999;184:297–315

**Figures**

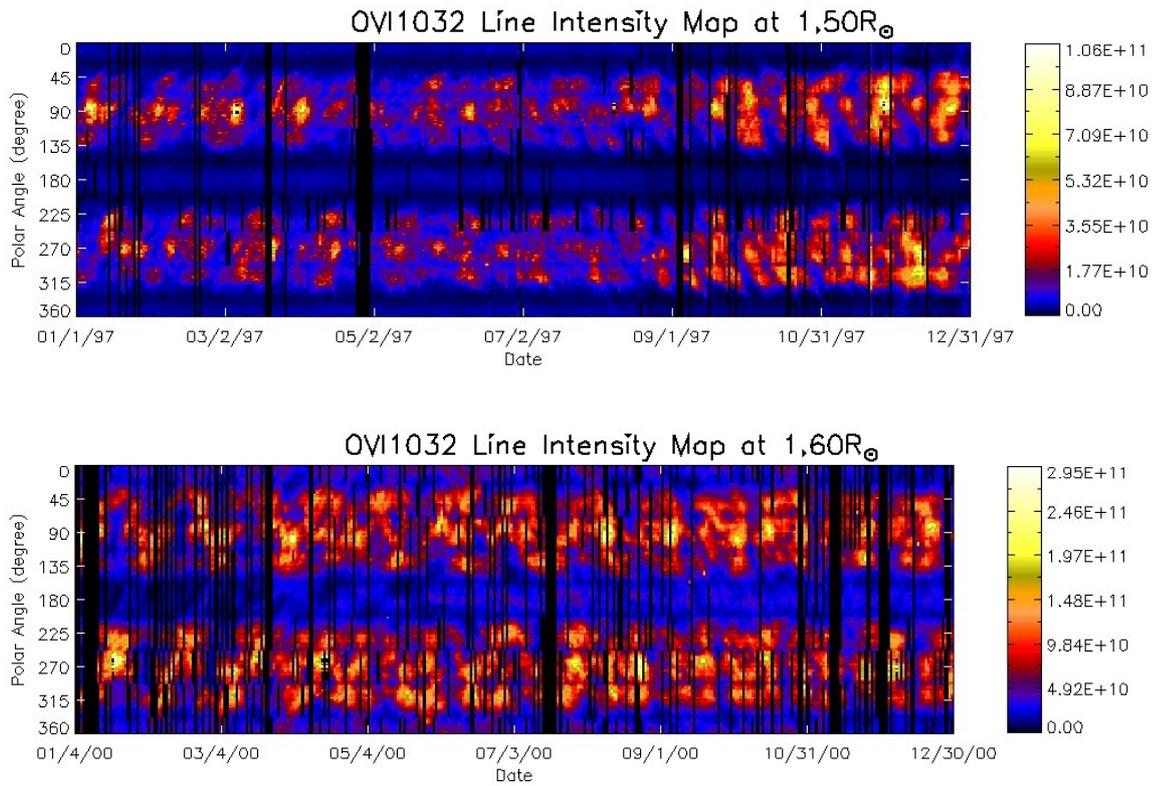

Fig. 1: Upper panel: O VI 1032 intensity synoptic map at 1.5 $R_\odot$ for year 1997 (solar minimum). Lower panel: O VI 1032 intensity synoptic map at 1.6 $R_\odot$ for year 2000 (solar maximum). Intensities are measured in units of photons cm$^{-2}$ s$^{-1}$ sr$^{-1}$. Position angles, measured counterclockwise (i.e., N-E-S-W-N) from the north pole, increase from top to bottom and cover all latitudes from 0° to 360°.

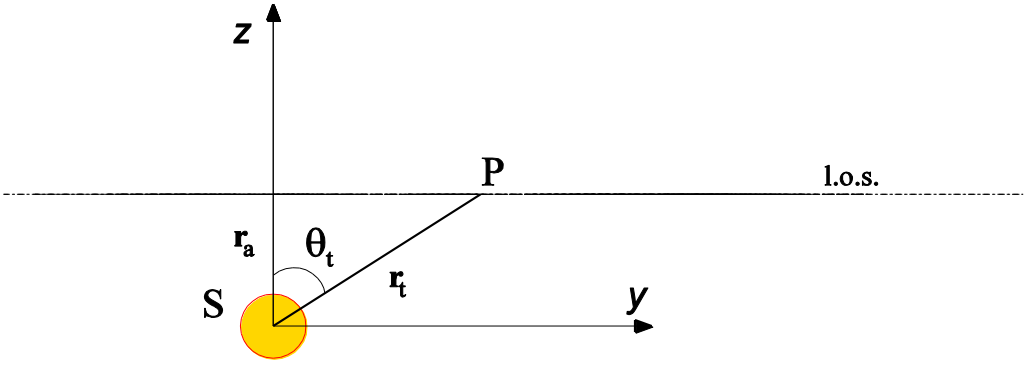

Fig. 2: Plot showing the geometrical model. See the text for details.

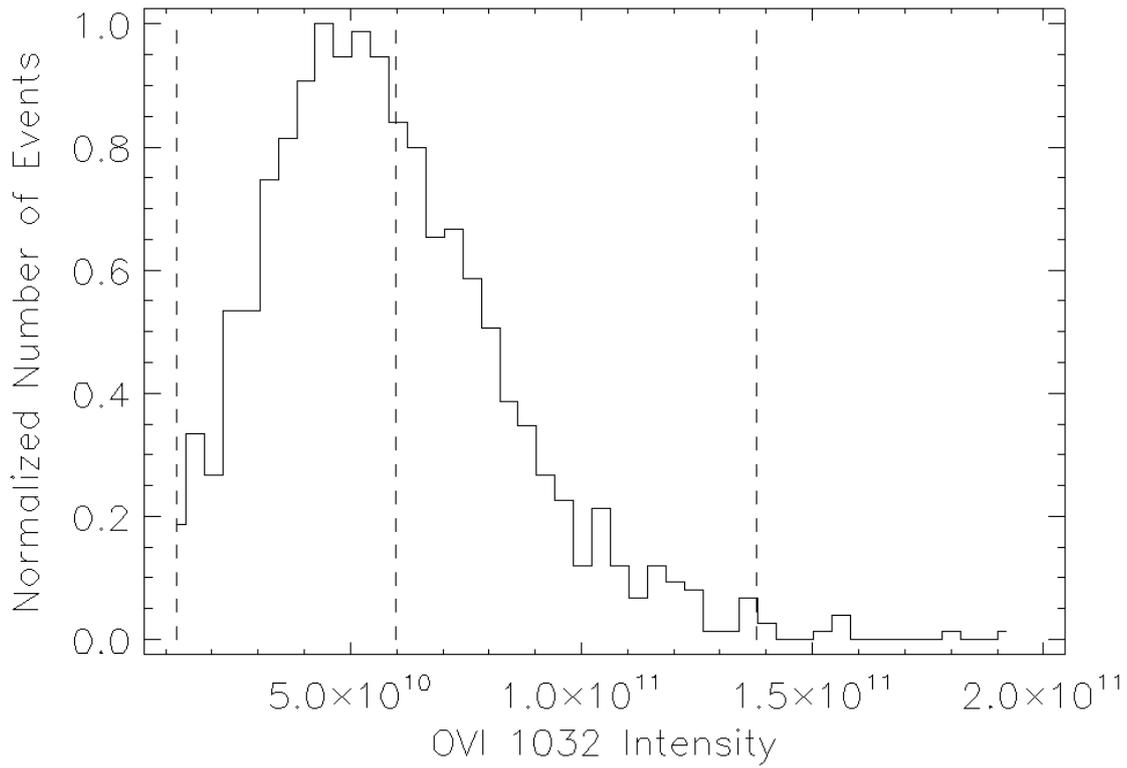

Fig. 3: O VI 1032 intensity distribution at 1.6 $R_\odot$. The dashed lines show the minimum and maximum expected intensity. The minimum intensity is assumed as typical of the background corona. The maximum intensity, computed as the average intensity plus 3σ, is the expected intensity of a bright streamer feature.

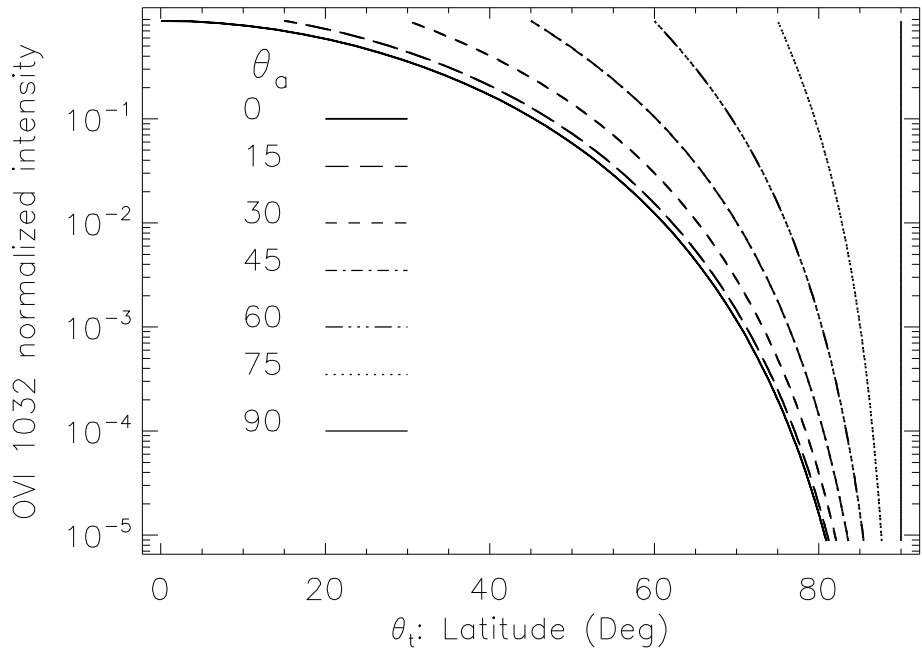

Fig. 4: Expected O VI 1032 intensity profile as a function of the true latitude, $\theta_t$, for different apparent latitudes, $\theta_a$.

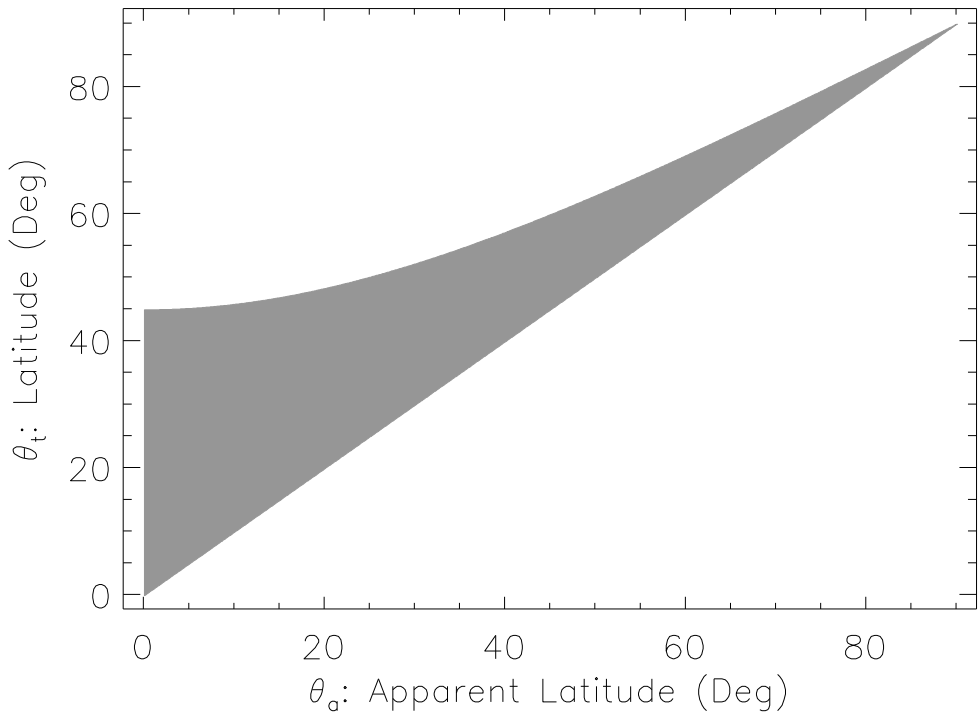

Fig. 5: Acceptable range of true latitude as a function of apparent latitude, in the dark region, a bright feature out of the plane of the sky can dominate the emission of the background corona.